# RNA Dynamics from Experimental and Computational Approaches

G. Bussi[1,*], M. Bonomi[2,*], P. Gkeka[3,*], M. Sattler[4,*], H. M. Al-Hashimi[5], P. Auffinger[6], M. Duca[7], Y. Foricher[8], D. Incarnato[9], A. N. Jones[10], S. Kirmizialtin[11], M. Krepl[12], M. Orozco[13], G. Palermo[14], S. Pasquali[15], L. Salmon[16], H. Schwalbe[17], E. Westhof[18], M. Zacharias[19]

[1]Scuola Internazionale Superiore di Studi Avanzati (SISSA), Via Bonomea 265, 34136 Trieste, Italy

[2]Institut Pasteur, Université Paris Cité, CNRS UMR 3528, Computational Structural Biology Unit, Paris, France

[3]Integrated Drug Discovery, Molecular Design Sciences, Sanofi, Vitry-sur-Seine, France

[4]Technical University of Munich and Helmholtz Munich, Germany

[5]Department of Biochemistry and Molecular Biophysics, Columbia University, New York, NY, USA

[6]Université de Strasbourg, Architecture et Réactivité de l'ARN, Institut de Biologie Moléculaire et Cellulaire du CNRS, 2 Allée Konrad Roentgen, 67084 Strasbourg, France

[7]Université Côte d'Azur, CNRS, Institute of Chemistry of Nice, Nice, France

[8]Integrated Drug Discovery, Small Molecules Medicinal Chemistry, Sanofi, Vitry-sur-Seine, France

[9]Department of Molecular Genetics, Groningen Biomolecular Sciences and Biotechnology Institute (GBB), University of Groningen, Groningen, the Netherlands

[10]Department of Chemistry, New York University, New York, New York, USA

[11]Chemistry Program, Science Division, New York University, Abu Dhabi, United Arab Emirates. Department of Chemistry, New York University, USA

[12]Institute of Biophysics of the Czech Academy of Sciences, Kralovopolska 135, Brno 612 00, Czech Republic

[13]Institute for Research in Biomedicine (IRB Barcelona). The Barcelona Institute of Science and Technology, Barcelona, Spain. Department of Biochemistry and Biomedicine. University of Barcelona, Barcelona, Spain

[14]Department of Bioengineering, The University of California, Riverside, California, USA. Department of Chemistry, The University of California, Riverside, California, USA

[15]Laboratoire Biologie Fonctionnelle et Adaptative, CNRS UMR 8251 INSERM ERL 1133, Université Paris Cité, 35 rue Hélène Brion, 75013 Paris, France

[16]Centre de RMN à Très Hauts Champs, UMR 5082 (CNRS, École Normale Supérieure de Lyon, Université Claude Bernard Lyon 1), University of Lyon, Villeurbanne 69100, France

[17]Institute for Organic Chemistry and Chemical Biology, Center for Biomolecular Magnetic Resonance, Goethe-University Frankfurt, 60438 Frankfurt/Main, Germany.

[18]Architecture et Réactivité de l'ARN, Université de Strasbourg, Institut de biologie moléculaire et cellulaire du CNRS, F-67084 Strasbourg, France

[19]Physics Department and Center of Protein Assemblies, Technical University of Munich, Germany

*Corresponding authors: bussi@sissa.it, mbonomi@pasteur.fr, Paraskevi.Gkeka@sanofi.com, michael.sattler@helmholtz-munich.de

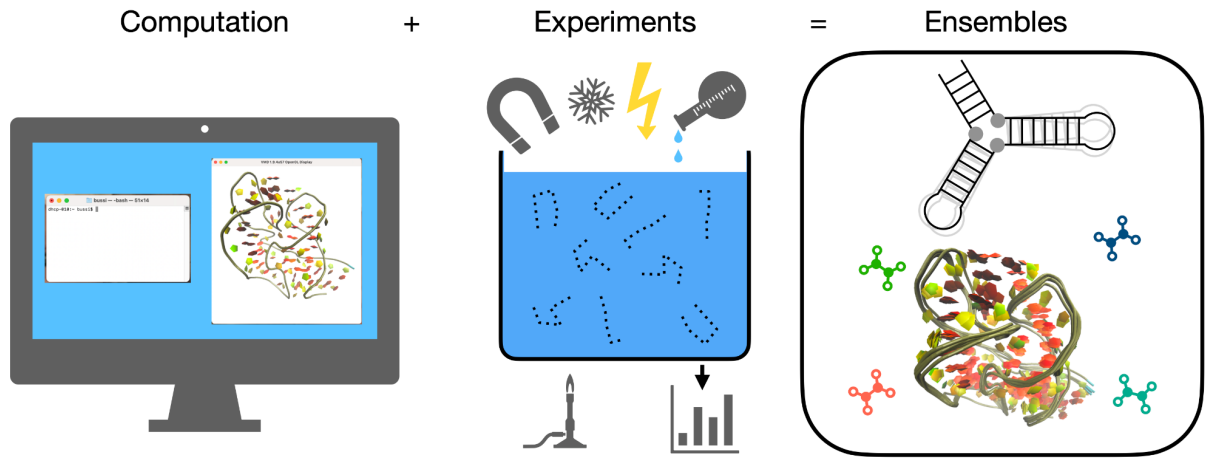

**Understanding RNA dynamics from combining molecular modeling and experiment.** Computations provide a detailed characterization of RNA structure using physics and chemistry-based models, which might be inaccurate (left panel). Data obtained using bulk techniques such as NMR, cryo-EM, X-ray scattering, chemical probing, or melting experiments provide average and often sparse information (central panel). The combination of simulations and experiments enables accurate reconstruction of structural ensembles, both at the secondary and tertiary structure level, which are key for studying the interaction of RNAs with other molecules (right panel).

**Abstract**

Ribonucleic acids (RNA) are unique in that they can store genetic information, replicate and perform catalysis. Importantly, RNA molecules are highly dynamic, and thus determining the ensemble of conformations that they populate is crucial not only to elucidate their biological functions, but also for their potential use as therapeutic targets. Computational and experimental techniques provide complementary views on RNA dynamics, and their integration is fundamental to improve the accuracy of computations and the resolution of experiments. Recent exciting developments in this field, were discussed at the CECAM workshop "RNA dynamics from experimental and computational approaches", in Paris, June 26-28, 2023. This report outlines key 'take-home' messages that emerged during this workshop from the presentations and discussions.

**Introduction**

Ribonucleic acid (RNA) molecules play fundamental roles in life sciences. RNAs are unique in that they can store genetic information, replicate and perform catalysis. In the cell, non-coding RNAs are pivotal for regulating gene expression at various stages [1]. Regulatory RNAs are hence emerging as promising therapeutic targets for either small molecules [2] or, for complementary, natural or artificial, antisense oligonucleotides [3]. Importantly, RNA molecules are highly dynamic, and thus determining the ensemble of conformations that they populate is crucial not only to elucidate their biological functions, but also for their potential use as therapeutic targets [4]. Computational techniques based on molecular dynamics (MD) simulations enable these structural ensembles to be reconstructed at atomistic resolution [5]. However, the accuracy of the employed energy models is limited, and thus experimental data are required to validate or refine the structural ensembles determined solely by MD [6]. From a complementary perspective, experimental techniques, such as nuclear magnetic resonance (NMR) spectroscopy, cryo-electron microscopy (cryo-EM), and, at a different scale, small/wide-angle X-ray (and/or neutron) scattering (SAXS/WAXS/SANS/WANS) and chemical probing experiments, can be used to provide information about RNA structural ensembles. However, in most cases these techniques report structural information that is averaged over the ensemble of conformations, often at limited resolution. Thus, integration of these experimental data with accurate physico-chemical models via advanced modeling tools is important.

Over the past decades, a variety of methods that integrate MD simulations and experiments have emerged [7]. Strong interactions between theoretical and experimental groups are fundamental to further advance the development of these approaches, to obtain a comprehensive, high-resolution understanding of RNA dynamics, and to ultimately create novel opportunities to develop therapeutic approaches targeting RNAs, also by exploiting the conformational dynamics for selective targeting of specific RNA states.

In light of these recent exciting developments, the CECAM workshop "*RNA dynamics from experimental and computational approaches*" was organized in Paris, on June 26-28, 2023, by Massimiliano Bonomi (Institut Pasteur - CNRS), Giovanni Bussi (SISSA), Paraskevi Gkeka (Sanofi), and Michael Sattler (Technical University of Munich and Helmholtz Munich). This workshop gathered scientists from academia and industry with selected contributions from participants to discuss state-of-the-art computational methods, their integration with experimental data, and applications to RNA structure, dynamics, and function. This report outlines the key 'take-home' messages that emerged during this workshop from the presentations and discussions.

**Molecular dynamics simulations and RNA structure prediction**

A number of sessions were dedicated to the application of **molecular dynamics simulations** to the characterization of the structural dynamics of RNA and RNA-protein complexes.

One of the key challenges in RNA modeling is the description of a given molecular system using a sufficiently accurate force field while simulating over physically relevant time scales. In this context and starting from DNA systems, Modesto Orozco (IRB Barcelona) illustrated how atomistic MD simulations allow us to describe structural and functional properties of DNA molecules with an accuracy comparable to experimental techniques, and discussed in which cases atomistic simulations can be used to parametrize coarse-grained and mesoscopic models. Achieving similar accuracy for RNA can be considered as a stepwise process, involving the simulation of non-natural nucleic acids with especially complex conformational flexibility, DNA·RNA hybrids, and finally pure RNA structures, for which improvements in the existing force-fields are needed to reproduce non-canonically paired regions. While accurate force fields for RNA are being developed, coarse-grained models can be directly trained on experimental data using an inverse-Boltzmann approach.

Miroslav Krepl (Institute of Biophysics, Czech Academy of Sciences) discussed how single-stranded RNAs (ssRNAs) are recognized by RNA binding proteins. The structures of many such complexes have been characterized using experimental techniques, which provide a static, ensemble-averaged picture of the complex. However, many regions at the protein-RNA interface are characterized by active competition between dynamic local substates rather than rigid interactions. Moreover, static structures cannot be used to infer details of the binding process. The Krepl group recently managed to decisively move beyond the confines of the bound state and simulated for the first time the spontaneous binding of short RNA sequences to RNA domains of HuR and SRSF1 proteins [8]. Aggregate ensembles of unbiased simulations covering the millisecond time scale allowed dozens of spontaneous binding events to be observed and provided information on binding mechanisms at atomistic resolution. To study spontaneous protein-RNA binding, a dedicated force-field modification was developed, where intramolecular van-der-Waals interactions were scaled. This new modification was also shown to improve the description of tightly bound complexes and can be widely applied in simulations of ssRNAs interacting with proteins.

In computational studies of RNA and DNA, the sequence-dependent deformability of helical segments is often described by harmonic stiffness models. Typically, unimodal models are employed based on a single stiffness matrix that can be obtained from the pre-computed stiffness of short double-stranded segments. Martin Zacharias (Technical University of Munich) presented a new multi-modal approach that combines the harmonic model with multiple meta-stable substates in DNA [9]. Possible extensions of this model to efficiently describe the flexibility of non-canonical structural motifs in RNA were discussed during the workshop. These approaches are mostly directed at studying local fluctuations in nucleic acids. The folding of large RNA-containing complexes is a complex process that is stabilized not only by base pairing, but also by other structural motifs such as metal-binding sites or loop-loop contacts. One recurring motif is the tetraloop:tetraloop-receptor interaction with a GNRA tetraloop contacting the minor groove of a target RNA receptor structure. Martin Zacharias presented advanced sampling free-energy simulations to study the absolute binding free energy of such interactions to elucidate the contribution to tertiary interactions in larger RNAs. Good agreement with available experimental data could be achieved for a model system and the influence of non-canonical base-pairs in the receptor could be characterized. This method also allowed estimating contributions from RNA conformational changes and deformability to binding that are essential for an in-depth understanding of the motif contribution to RNA tertiary structure formation and stability.

The structural investigation of RNAs in a physiological environment presents various challenges from an experimental perspective. The high sensitivity of RNA to salt necessitates replicating similar conditions *in vitro*. However, structural methods that provide atomic resolution information require specific sample conditions, making it challenging to establish meaningful comparisons between *in vitro* and *in vivo* scenarios. This underscores the necessity of studying RNA dynamics using solution experimental techniques. While achieving atomic-scale resolution solely through solution methods remains a substantial obstacle, the integration of molecular simulations with such studies unveils an efficient integrative approach. Serdal Kirmizialtin (New York University, Abu Dhabi) showed how wide-angle X-ray scattering (WAXS) combined with MD simulations can provide information about nucleic acids at atomic resolution. This integrative approach allows the exploration of RNA conformational ensembles across a spectrum of salt conditions and RNA topologies [10]. Furthermore, comparing WAXS signals with computed profiles from MD simulations allows overcoming the imbalance in empirical energy potentials, and has the potential of leading to more accurate RNA force fields.

RNAs are flexible molecules that can adopt a variety of conformations that are often separated by high interconversion free-energy barriers. These conformations may adopt different secondary or tertiary structures. Because transitioning between different states is a rare event, traditional simulation methods are limited. To address such challenging systems, Pasquali (Université Paris Cité) and collaborators have adopted the discrete path sampling approach based on geometric optimization and complemented it with Hamiltonian replica exchange MD simulations [11]. This combined approach enabled characterization at atomistic resolution of the full conformational landscape for systems up to a few dozen nucleotides, highlighting the presence of alternative states along with their transition pathways. For the method to be applicable to larger systems and different external conditions such as pH, Pasquali's group developed a coarse-grained (CG) model, named HiRE-RNA. Using a combination of this CG model and a fast proton exchange Monte Carlo scheme, the group has performed constant pH simulations and determined how changes in pH modulate the population of different RNA conformations.

In recent years, CRISPR-based technologies have become transformative tools for genome editing with cutting-edge impact in basic and applied sciences and translation. Giulia Palermo (University of California, Riverside) discussed how multiscale computational approaches have been instrumental in clarifying the molecular basis of CRISPR systems [12]. Enhanced-sampling MD and the Anton-2 massively parallel supercomputer were used to determine the formation of a catalytically active state. These computational predictions have been confirmed by single-molecule fluorescence resonance energy transfer experiments and by the recent determination of the active state with cryo-EM. By applying network models derived from graph theory, a mechanism of allosteric regulation that transfers the information of DNA binding to the catalytic sites for cleavages has been proposed. A mixed quantum mechanics/molecular mechanics approach was used in combination with enhanced-sampling MD to establish the catalytic mechanism of DNA cleavage and the critical role of metal ions, providing information useful for improving the catalytic efficiency and the specificity of CRISPR-Cas9. Finally, by using multi-microsecond MD simulations, a DNA-induced activation mechanism was studied in the Cas12a enzyme, clarifying the critical mechanistic aspects of nucleic acid detection. Overall, these studies contribute to addressing the mechanistic function of CRISPR-Cas systems, providing information that is critical for novel engineering and that could accelerate the development of improved genome editing tools for biomedical applications.

Despite the fact that cryo-EM studies on RNA systems have been reaching unprecedented resolution, the models derived from these data sometimes contain imprecisions in the assignment and placement of ions and in the modeling of important structural motifs. Moreover, our understanding of the forces involved in structural stability and functional dynamics is incomplete, especially regarding the weakest of these forces, such as the oxygen-π interactions that are found in UNCG/GNRA tetraloops and are notoriously challenging for MD simulations. Quantum mechanical calculations demonstrate that these oxygen-π contacts are weak and are imposed by the tight stereochemical constraints present in these motifs and not by strong interaction forces [13]. Calculations from Pascal Auffinger (Université de Strasbourg) and co-workers suggest that recalibrating the van der Waals radii of the atoms composing the nucleobase aromatic rings might contribute to improving the quality of the modeling of these RNA structures and of related motifs. Auffinger also advocates for a better understanding of the solvent structure of these biomolecules given that, even with the best resolutions currently available, misassignments of solvent molecules are made and affect the perception of the role of $Mg^{2+}$ and $K^+$ ions, ultimately limiting the accuracy of MD simulations that employ these models.

3D structure modeling of RNA molecules traces back to the late 1960s, and, since then, numerous computational approaches have emerged for predicting RNA 3D structures. RNA-Puzzles developed by Eric Westhof (IBMC-CNRS, Université de Strasbourg) and co-workers is a collaborative effort dedicated to advancing and improving RNA 3D structure prediction [14]. With the agreement of crystallographers, RNA structures are predicted by different groups before the publication of crystal structures and compared systematically afterward. Stimulated by the success of deep-learning methods for protein structure prediction, such as AlphaFold, novel artificial intelligence approaches have been designed to solve the problem of RNA 3D structure prediction. However, eliminating redundancy between training and test data is not trivial and some approaches have shown overfitting results. Therefore, blind and unbiased evaluations of all prediction tools based on established comparison metrics are a requirement. Eric Westhof presented a dedicated website (http://www.rnapuzzles.org/) that gathers the systematic protocols and parameters used for comparing models and crystal structures, all the data, analysis of the assessments, and related publications. Up to now, 40 RNA sequences with structures determined experimentally by X-ray or cryo-EM have been predicted by many groups from several countries.

After comparison with the solved structures, many of the predictions demonstrated good accuracy in global folds, but poor accuracy in local structure, which is key for understanding molecular recognition.

**Combining experimental approaches and MD simulations to study RNA structure and dynamics**

A number of presentations highlighted the unique roles of **NMR spectroscopy** techniques to probe RNA structural dynamics and demonstrated their important roles for biological function.

Knowledge of RNA structural dynamics is essential for quantitatively modeling the formation of the RNA-protein complexes and the cellular processes they drive. These interactions nearly universally require changes in the RNA conformation. As a result, binding affinity and cellular activity depend not only on the strength of the contacts but also on the inherent propensities to form binding-competent conformational states. However, conceptual and technological limitations have hindered the ability to quantitatively measure how conformational propensities impact cellular activity. Hashim Al-Hashimi (Columbia University) described a study that used mutations to systematically alter the propensities for forming the protein-bound conformation of HIV-1 TAR RNA [15]. These conformational propensities quantitatively predicted the binding affinities of TAR to the RNA-binding region of the Tat protein and predicted the extent of HIV-1 Tat-dependent transactivation in cells. These results establish the role of ensemble-based conformational propensities in cellular activity and highlight an example of a cellular process driven by an exceptionally rare and short-lived RNA conformational state.

Loic Salmon (University of Lyon) presented novel methodologies to combine molecular modeling with NMR data to explain the dynamics of complex RNAs. Different types of experimentally accessible solution data were discussed, in particular NMR residual dipolar couplings and their combination using orthogonal alignment to access key RNA motions at timescales up to the millisecond [16]. Different types of molecular modeling were tested to sample as extensively as possible RNA conformational landscapes up to this timescale. NMR data were used to reweight this initial ensemble to obtain a more accurate description of RNA motions. The work proposed an initial model of highly flexible RNA stretches and highlighted the benefits of combining atomic computational models and detailed experimental data for such systems.

Harald Schwalbe (Goethe University Frankfurt) presented a series of NMR investigations of RNAs. He first discussed how RNA structure is inherently dynamic. A joint NMR/MD study in collaboration with the groups of Lindorff-Larsen and Blackledge indicates that a thermodynamically highly stable CUUG RNA tetraloop undergoes substantial temperature-dependent dynamics. MD simulations, refined by the application of a plethora of NMR restraints, are required to characterize the structural ensemble even for such small systems [17]. In addition to the inherent structural dynamics of RNA elements, RNA molecules undergo post-transcriptional modifications. Schwalbe discussed the impact of m6A-modification on the SL2_3 RNA element located in the 5'-genomic end of the genome of SARS-CoV-2. This element contains a DRACH element as m6A target side and builds in addition the transcriptional-regulatory-sequence (TRS)-leader sequence for the discontinuous transcription leading to the expression of the viral structural proteins. m6A does not affect the structure of the RNA element but leads to a destabilization of the interaction of the TRS-leader and the TRS-body sequence. Together with the group of Wöhnert and Duchardt-Ferner, Schwalbe and Wacker determined the structures of the regulatory elements of SARS-CoV-2, which revealed a high number of unexpected base pairs, for which high-resolution NMR structure determination was shown to be essential.

One session was dedicated to the use of **chemical probing experiments** to elucidate RNA secondary structure and dynamics.

In chemical probing assays, small reactive molecules irreversibly modify single-stranded or flexible nucleotides. These modifications are assessed through reverse transcription (RT) reactions, where the reverse transcriptase either stalls (RT-stop) or introduces a mutation (RT-MaP) at the site of the modified nucleotide. Ideally, modifications in RT-stop experiments reflect a single-hit kinetic mechanism, where one modification occurs per molecule of RNA. In RT-MaP experiments, multi-hit kinetics can be observed, but experiments are carried out such that modifications are sparse.

Alisha Jones (New York University) discussed recent work where atomistic MD simulations suggest that binding of RNA by 1-methyl-7-nitroisatoic (1M7) is affected by cooperative effects, leading to an observed reactivity that is dependent on the concentration of the chemical probe [18]. In her lab, this effect was then assessed with numerous chemical probes. As the concentration of a chemical probe increases, the binding and modification of a nucleotide can locally modulate the RNA structure, resulting in an increase or decrease of chemical probe binding and reactivity in the surrounding nucleotides. This cooperative effect depends on both chemical probe concentration and size, and interestingly, occurs at concentrations intended to ensure single-hit and sparse multi-hit kinetic schemes. A consequence of this effect is that 2D structures, which are often built using reactivity profiles, can be mispredicted. This work highlights the importance of optimizing chemical probing experiments to predict RNA secondary structures accurately.

Chemical probing experiments are powerful means to investigate RNA structures in living cells. For decades, these experiments have been used to determine a single RNA structure, completely disregarding the conformationally heterogeneous nature of RNA molecules. Indeed, since RNAs often populate an ensemble of coexisting structures, the reactivities derived from chemical probing experiments represent a weighted average over the structural ensemble. Methods that enable the readout of multiple sites of chemical modification sites on a RNA molecule are collectively referred to as mutational profiling (MaP) methods. Danny Incarnato (University of Groningen) reviewed recent computational tools that have been transformative in providing increased resolution of RNA chemical probing data, to the point of enabling the deconvolution of alternative structures populating ensembles of secondary structures *in vivo* conditions [19]. For these applications, the choice of the chemical probe is critical in relation to the size of the structurally heterogeneous region within the target RNA. For instance, DMS (dimethyl sulfate) has a higher signal-to-noise ratio as compared to conventional SHAPE (Selective 2' Hydroxyl Acylation analyzed by Primer Extension) reagents, but, traditionally, it can only probe ~50% of the bases in the transcriptome (A and C). In this perspective, improved SHAPE compounds such as 2-aminopyridine-3-carboxylic acid imidazolide (2A3) provide significantly higher signal-to-noise ratio, further enabling efficient ensemble deconvolution.

**Pharmaceutical applications**

Targeting RNA with small molecules is a booming area of drug discovery that can potentially provide solutions for patients with unmet medical needs. Yann Foricher (Sanofi) provided an overview of recent developments and promising efforts in this area. Since the discovery of messenger RNA in 1961, knowledge about RNA biology and different RNA types as well as their functional role has been increasing exponentially. This paved the way to design strategies to target RNA or to use natural or artificial nucleic acids as therapeutic modalities. Ribosomal antibiotics were the first small molecules to be developed and approved by drug agencies. Outside antibiotic drugs, RNA has been a challenging target for small molecules for many years, due to lack of selectivity. Over the past 15 years, understanding of *in vitro* and *in vivo* RNA biology, progress in structural biology, and several technical developments have paved the way to discover selective and potent drug-like molecules in a rational manner. Biotech and pharmaceutical companies are pursuing various approaches to achieve such a goal focusing on four different areas: RNA translation, RNA splicing modulation, direct RNA targeting, and epitranscriptomics. Recently, following the impressive work carried out by Roche Pharma, PTC and Spinal Muscular Atrophy (SMA) Foundation, risdiplam was approved by FDA for SMA Disease based on its efficacy, selectivity, and safety clinical data. This clearly represents a key milestone for RNA-small molecule drug discovery [2,20]. Despite some remaining challenges in these new R&D efforts for small molecules, especially in RNA splicing modulation and direct RNA targeting, the first molecules designed by a rational approach were expected to soon enter into clinical trials and indeed Skyhawk Therapeutics recently announced SKY-0515 small molecule candidate targeting Huntington's disease (HD) entering Phase I.

Maria Duca (Université Côte d'Azur) focused on the development of new strategies for targeting non-coding RNAs using small molecules. Using both structure-based design and high-throughput screening, the Duca lab identified new selective binders against microRNAs precursors, such as pre-miRNAs, that inhibit their maturation toward mature oncogenic miRNAs [21]. Such binders target the inhibition of proliferation; their effect was demonstrated in cancer cell lines as well as in patient-derived cancer cells. To illustrate the potential of this approach, examples of compounds designed to inhibit RNA-RNA interactions have been shown. Those ligands could find application in the targeting of bacterial non-coding RNAs toward innovative antimicrobial therapies.

**Discussion**

The CECAM workshop provided a great opportunity to discuss the state-of-the-art and future perspectives of understanding and leveraging RNA dynamics. An animated discussion session evolved around several questions that emerged during the scientific presentations.

*How can one properly compare and integrate experimental data and computational simulations?* Simulations can produce a vast amount of structural and dynamic data, which in many cases can be directly validated against raw experimental information. Compared to a few decades ago, the amount of raw data deposited on the web has significantly increased, enabling computational groups to directly use this information. At the same time, many modeling tools are becoming more and more accessible to the experimental community, thanks to their comprehensive documentation and increased usability. However, in both computational and experimental areas, significant experience is still required to judge the reliability/quality of the produced information, especially when advanced computational or experimental techniques are used. Direct collaborations between experimental and computational groups are therefore recommended to make the best out of the techniques developed by both communities.

*How can the accuracy of current force fields used in molecular dynamics simulations be improved?* Whereas current prediction tools can model individual RNA structures from first principles at a reasonable accuracy, MD force fields still have some limitations in accurately describing RNA dynamics, for example, due to a poor treatment of the interaction between divalent cations and RNA and the lack of polarizability effects. As a matter of fact, for flexible RNA oligomers, it is still difficult to model structural ensembles that are in agreement with experiments and avoid overfitting. These limitations make the integration with experimental data even more important. Experiments can be used to refine ensembles and provide fundamental information for improving MD force fields.

*How can artificial intelligence approaches be leveraged?* Artificial intelligence and in particular deep learning methods have revolutionized the field of protein structure prediction and are increasingly being explored for their potential to reconstruct protein dynamics as well. Applications to RNA have been relatively scarce so far. This might be a consequence of the significantly smaller number of available structures that can be used to train these tools [22]. In addition, the fact that RNA sequences are composed of a more limited and homogeneous alphabet makes extracting useful co-evolutionary information more difficult. Observing the developments of this field over the coming years will be intriguing.

*How well do we understand and can we predict RNA-protein interactions?* The difficulties discussed above are amplified in the case of RNA-protein complexes, where the number of interactions that should be properly modeled is increased dramatically. This is an area of research where artificial intelligence structure predictors have just started making their contribution, as demonstrated by the recent release of RoseTTAFoldNA [23] and the announcement of the forthcoming version of AlphaFold.

*How to assess the biological and pharmaceutical relevance of RNA structure and dynamics?* Experimental data suggest that RNA structures are highly dynamic, and therefore an individual structure is in most cases not sufficient to explain measurements performed in solution. However, this observation is *per se* not suggestive of the relevance of RNA dynamics *in vivo*. In this sense, biophysical studies performed *in vitro*, possibly complemented with advanced modeling techniques, can suggest the coexistence of multiple structures, and help generate hypotheses regarding the possible mechanisms of action *in vivo*. These hypotheses could be further validated by performing biological assays including mutations designed to alter RNA dynamics in a controlled manner. Similarly, exploiting RNA dynamics in pharmaceutical applications has unique potential for innovative therapeutic approaches, but still has to be fully demonstrated. It is however very reasonable to expect that standard rigid-docking approaches might fail in characterizing the interaction of flexible RNA with drug-like molecules.

*Future perspectives.* The CECAM meeting organized in Paris has been very successful in bringing together a heterogeneous community composed of computational and experimental scientists with strong complementary expertise in different techniques to predict and measure RNA dynamics. Established researchers and early-career scientists presented their work in the various sessions, and given the informal setting, early-career scientists had ample opportunities to directly interact with more experienced colleagues in the field. The meeting has attracted a lot of attention from the community. Unfortunately, given the limited capability of the venue, only a fraction of the applicants could be admitted. To mitigate this problem, the meeting was streamed to be accessible to all applicants, thus significantly increasing its outreach. A number of new collaborations were initiated during the meeting, and some of the participants are planning to organize events in the same spirit in different geographical locations, maximizing inclusivity and reaching out to a wider scientific community. For example, this meeting brought together for the first time the community working on different aspects of RNA structure and dynamics in the Paris area. After the workshop, this group of people has already met twice to discuss more in detail the research carried out in different labs and to foster new collaborations.

**Acknowledgments**

CECAM-FR-MOSER, CECAM-IT-SISSA-SNS, Schrödinger Gmbh, and Sanofi are acknowledged for funding the CECAM workshop "RNA dynamics from experimental and computational approaches" organized in Paris, on June 26-28, 2023.

## Competing interest

P. Gkeka is or was a Sanofi employee and may own stocks in Sanofi.